\documentclass[letterpaper]{article}
\usepackage{aaai2027}
\usepackage[hyphens]{url}
\usepackage{graphicx}
\usepackage{natbib}
\usepackage{caption}
\usepackage{algorithm}
\usepackage{algorithmic}
\usepackage{booktabs}
\usepackage{amsmath}
\usepackage{amssymb}
\usepackage{bm}
\usepackage{multirow}
\newtheorem{proposition}{Proposition}

\newcommand{\Vpre}{V_{\mathrm{pre}}}
\newcommand{\Vpost}{V_{\mathrm{post}}}
\newcommand{\Iaqdpc}{\mbox{IA-VQC-DPC}}
\newcommand{\degK}{\,\mathrm{K}}

\title{Who Earns the Safety? Intervention-Aware Quantum Predictive Control with Safety Attribution}

\author{
    Yifan Wang
}
\affiliations{
    Department of Mechanical Engineering, McGill University\\
    Montr\'eal, QC H3A 2T7, Canada\\
    yifan.wang18@mail.mcgill.ca
}

\begin{document}
\maketitle

\begin{abstract}
Hard safety filters are increasingly placed downstream of learned controllers to guarantee constraint satisfaction at run time. Yet a filtered controller that never violates a constraint may still have learned nothing about safety: the filter can silently repair an incompetent upstream policy, so that post-filter success measures the \emph{filter}, not the \emph{policy}. We argue that safe policy learning should ask \emph{who earns the safety}---the policy or its protective layers---and we make this question measurable. We introduce \textbf{Intervention-Aware Variational Quantum Differentiable Predictive Control} ($\Iaqdpc$), which (i) trains a \emph{compact} variational quantum circuit (VQC) policy under a primal--dual \emph{intervention budget} that penalizes reliance on a differentiable Control-Barrier-Function (CBF) projection, and (ii) is evaluated with a \emph{safety-attribution} protocol that decomposes the executed-trajectory correction into a CBF term and a deployment \emph{runtime-guard} term, and stress-tests the policy with \emph{guard-off} evaluation. On closed-loop, high-fidelity \mbox{BOPTEST} building-control emulators ($5$ seeds, $60$ episodes per method), intervention-aware training significantly lowers the quantum policy's raw pre-filter violation and total safety-layer reliance (both $p<10^{-4}$) with no significant energy regression; at an equal ${\approx}400$-parameter budget the quantum policy is significantly safer and more comfortable than a matched classical policy. Guard-off evaluation confirms the improvement is \emph{policy-level} and exposes a valuable negative result: a learned differentiable energy head is only safe when paired with a distribution-aware runtime guard. The attribution protocol is general beyond quantum policies and buildings.
\end{abstract}

\section{Introduction}
Learning-based controllers are attractive for cyber-physical systems such as building heating, ventilation, and air-conditioning (HVAC), where model predictive control already delivers large comfort and energy gains \citep{drgona2020mpc}. To deploy such controllers safely, a common pattern wraps the learned policy with a \emph{hard safety filter}---for example a Control-Barrier-Function (CBF) projection or a predictive safety filter---that minimally edits each proposed action so the executed action provably respects state constraints \citep{ames2019cbf,wabersich2021safetyfilter,alshiekh2018shielding}. The filter is a guarantee at run time; it is also, we argue, an \emph{evaluation hazard}.

\begin{figure*}[t]
\centering
\includegraphics[width=\textwidth]{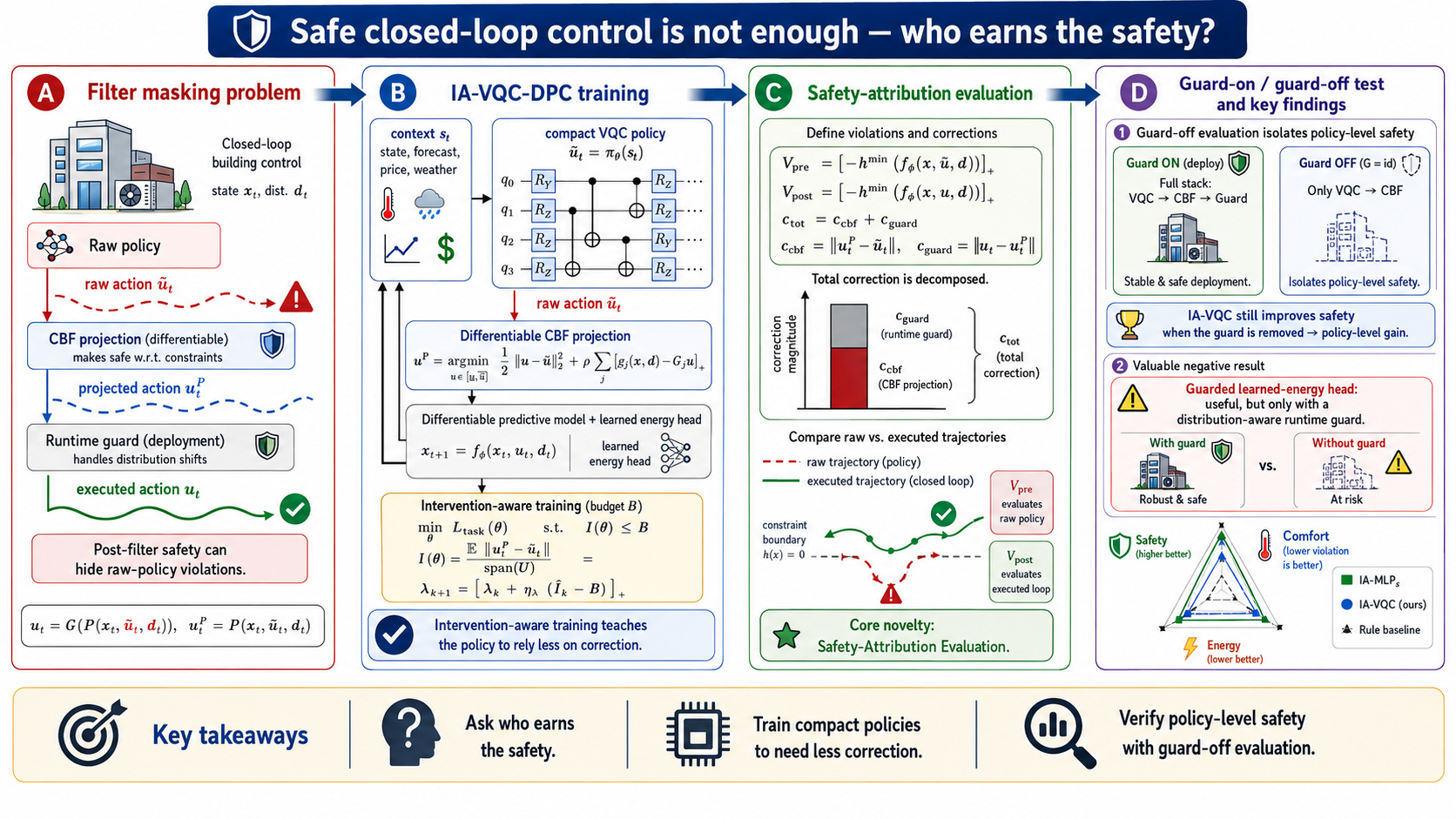}
\caption{\textbf{Who earns the safety?} \emph{(Left)} A compact variational quantum circuit (VQC) policy proposes an action that a differentiable Control-Barrier-Function (CBF) projection and a deployment runtime guard edit before execution, so a post-filter loop can look safe even when the raw policy is not. \emph{(Center)} Intervention-aware training adds a primal--dual \emph{intervention budget} that penalizes reliance on the safety layers, pushing the policy to be intrinsically close to safe. \emph{(Right)} Our \emph{safety-attribution} evaluation decomposes the executed correction into a CBF term and a runtime-guard term and adds a \emph{guard-off} stress test, so measured safety is credited to the policy that earned it rather than to the filter that masked it.}
\label{fig:teaser}
\end{figure*}

\paragraph{The filter-masking problem.}
If a learned policy repeatedly proposes unsafe actions and the filter repairs them, the closed loop still reports zero violations and possibly high reward. Standard metrics---post-filter constraint satisfaction and task return---then certify a policy that has internalized \emph{no} safe-control structure. The credit for safety belongs to the filter. This conflation is rarely measured: existing safe reinforcement learning (RL) and shielding methods \emph{ensure} safety by constraining or correcting actions \citep{garcia2015survey,achiam2017cpo,dalal2018safe,cheng2019safe}, but they typically do not ask whether the upstream policy has actually \emph{learned} to be safe, nor do they separate the policy's contribution from the filter's.

\paragraph{Why a compact quantum policy is a sharp test case.}
The masking problem is most acute for compact, expressivity-constrained policy classes, whose raw behavior is easiest for a filter to dominate. Variational quantum circuit (VQC) policies are precisely such a class: they are parameter-efficient and carry a periodic (Fourier) inductive bias from data re-uploading \citep{mitarai2018qcl,schuld2021fourier,jerbi2021parametrized}, but their value in realistic constrained control is unproven and easily overstated. A VQC wrapped by a strong filter can look competent while the classical projection does the work. Quantum policies therefore need intervention-aware evaluation \emph{more} than reward-only evaluation---and, conversely, they are an ideal probe for whether intervention-aware training actually teaches a small policy to be safe.

\paragraph{Contributions.}
We make the ``who earns the safety'' question both a training signal and an evaluation protocol (Fig.~\ref{fig:teaser}), and instantiate it with a compact quantum policy on high-fidelity building-control closed loops:
\begin{enumerate}\itemsep1pt
\item \textbf{Safety-attribution evaluation.} A closed-loop protocol that decomposes the executed correction into a \emph{CBF} term and a deployment \emph{runtime-guard} term, reports raw \emph{pre-filter} versus \emph{post-filter} violation, and adds a \emph{guard-off} stress test that isolates policy-level safety. This exposes that naive ``correction-norm'' comparisons are confounded by deployment guards (Sec.~\ref{sec:exp}).
\item \textbf{$\Iaqdpc$.} An intervention-aware differentiable-predictive-control objective that trains a VQC policy under a \emph{primal--dual intervention budget} on a differentiable CBF projection, so the policy is pushed to be intrinsically close to safe rather than to lean on correction (Sec.~\ref{sec:method}).
\item \textbf{Guarded learned-energy head (a valuable negative result).} We show a learned differentiable energy model improves the controller \emph{only} when paired with a distribution-aware runtime guard; without the guard it is exploited out of distribution, producing physically pathological closed loops (Sec.~\ref{sec:exp}).
\item \textbf{Evidence on high-fidelity emulators.} On BOPTEST v0.9.0 closed loops ($5$ seeds, $60$ episodes/method), $\Iaqdpc$ significantly reduces raw pre-filter violation and total safety-layer reliance ($p<10^{-4}$) with no significant energy regression, and at an equal ${\approx}400$-parameter budget is significantly safer and more comfortable than a matched classical policy; guard-off evaluation confirms the gain is policy-level.
\end{enumerate}
Together these turn ``who earns the safety'' from a slogan into a measurable, trainable property: a compact quantum policy is driven to earn its own safety, and the attribution and guard-off protocol makes that earned safety visible exactly where post-filter metrics hide it.

\section{Related Work}
\paragraph{Safe RL and shielding.}
Constrained-MDP and Lagrangian methods bound expected cost \citep{achiam2017cpo,stooke2020pid}; shielding and action projection enforce hard constraints by overriding unsafe actions \citep{alshiekh2018shielding,dalal2018safe,cheng2019safe}; surveys catalog the landscape \citep{garcia2015survey}. These works guarantee or improve safety but evaluate the \emph{executed} closed loop; they do not attribute safety between the policy and its corrective layers, which is our focus.

\paragraph{Safety filters and differentiable optimization.}
CBFs give forward-invariant safe sets \citep{ames2019cbf} and predictive safety filters minimally modify actions \citep{wabersich2021safetyfilter}. Differentiable optimization layers \citep{amos2017optnet,agrawal2019diffcvx} make such projections trainable end to end. We use a differentiable, slackened CBF projection but treat the \emph{intervention it induces} as a first-class training and evaluation signal.

\paragraph{Differentiable predictive control.}
DPC learns explicit control policies through differentiable closed-loop rollouts and is well suited to buildings \citep{drgona2022dpc,drgona2020mpc}. We extend DPC with an intervention-aware objective and a quantum policy class, and we are explicit that our building instantiation uses one-step differentiable model losses rather than a multi-step rollout.

\paragraph{Quantum policies.}
Data re-uploading VQCs are expressive parametric models with a Fourier structure \citep{mitarai2018qcl,benedetti2019pqc,schuld2021fourier}, and parametrized quantum policies have been used in RL \citep{jerbi2021parametrized,skolik2022gym}. Such policies are typically evaluated on small, weakly constrained tasks and without hard safety filters; we evaluate a compact VQC under hard filters on a realistic constrained benchmark, in the NISQ-relevant regime \citep{preskill2018nisq}.

\paragraph{Building benchmarks.}
BOPTEST \citep{blum2021boptest} and CityLearn \citep{vazquez2019citylearn} provide realistic, constrained building-control environments. We use BOPTEST v0.9.0 hydronic test cases as a realistic testbed, not as the methodological novelty.

\section{Problem Setup and Attribution Metrics}
\label{sec:setup}
We consider discrete-time closed-loop control of a building zone with thermal state $x_t\in\mathbb{R}^{n_x}$, control $u_t\in\mathcal{U}\subset\mathbb{R}^{n_u}$ (actuator commands), and exogenous disturbances $d_t$ (weather, occupancy, price). A predictive model $x_{t+1}=f_\phi(x_t,u_t,d_t)$ supports differentiable losses. Comfort defines a safe set $\mathcal{C}=\{x:h_j(x)\ge 0,\ j=1,\dots,m\}$ with $h_j$ the signed temperature-band margins.

\paragraph{Raw and executed actions.}
A policy proposes a \emph{raw} action $\tilde u_t=\pi_\theta(s_t)$ from a context $s_t$ (observations and short forecasts). Before execution it passes through two layers: a differentiable CBF projection $\mathsf{P}$ and a deployment \emph{runtime guard} $\mathsf{G}$,
\begin{equation}
u_t \;=\; \mathsf{G}\big(\mathsf{P}(x_t,\tilde u_t,d_t)\big), \qquad u^{\mathsf P}_t \;=\; \mathsf{P}(x_t,\tilde u_t,d_t).
\end{equation}
The CBF projection enforces the discrete barrier condition $h_j(f_\phi(x,u,d))\ge(1-\alpha_j)h_j(x)$ through a slackened, box-constrained quadratic program with an exact penalty,
\begin{equation}
u^{\mathsf P}=\!\!\arg\min_{u\in[\,\underline u,\overline u]}\ \tfrac12\|u-\tilde u\|_2^2+\rho\!\sum_{j}\big[g_j(x,d)-G_j u\big]_+,
\label{eq:proj}
\end{equation}
where $[\cdot]_+=\max(\cdot,0)$, $G_ju\ge g_j$ encodes the affine-in-$u$ barrier condition, and $\rho$ is a large penalty. The runtime guard $\mathsf G$ is a deployment-time, distribution-aware layer (a logged-safe action envelope plus a hydronic actuator-consistency rule) that prevents out-of-distribution actuator combinations that are within simple bounds yet physically pathological.

\paragraph{Attribution metrics.}
We separate \emph{who corrected what}. With $h^{\min}(x)=\min_j h_j(x)$, the \emph{raw pre-filter violation} and \emph{post-filter violation} are
\begin{equation}
\begin{aligned}
\Vpre&=\big[-h^{\min}\big(f_\phi(x,\tilde u,d)\big)\big]_+,\\
\Vpost&=\big[-h^{\min}\big(f_\phi(x,u,d)\big)\big]_+ .
\end{aligned}
\end{equation}
$\Vpre$ asks whether the \emph{raw} policy was safe; $\Vpost$ asks whether the \emph{executed} loop was safe. The raw action is edited in two \emph{sequential} stages---the CBF projection $\tilde u\!\to\!u^{\mathsf P}$ and the deployment runtime guard $u^{\mathsf P}\!\to\!u$---and we define the total safety-layer reliance as the sum of the two stage corrections,
\begin{equation}
c_{\mathrm{cbf}}:=\|u^{\mathsf P}-\tilde u\|,\quad c_{\mathrm{guard}}:=\|u-u^{\mathsf P}\|,\quad c_{\mathrm{tot}}:=c_{\mathrm{cbf}}+c_{\mathrm{guard}}.
\label{eq:decomp}
\end{equation}
Reporting $c_{\mathrm{cbf}}$ and $c_{\mathrm{guard}}$ separately is essential: a method can reduce $c_{\mathrm{tot}}$ merely by staying inside the deployment guard, which is \emph{not} evidence that it learned the CBF-relevant safe-control structure.

\section{$\Iaqdpc$}
\label{sec:method}

\paragraph{Compact quantum policy.}
The raw policy is a data re-uploading VQC on $n_q$ qubits. A classical encoder maps the context to angles $e(s)=\pi\tanh(W_e s+b_e)\in[-\pi,\pi]^{n_q}$; each of $L$ layers applies $\mathrm{RX}(\lambda_{l,q}e_q(s))$, then $\mathrm{RY}(\theta^{(1)}_{l,q})\,\mathrm{RZ}(\theta^{(2)}_{l,q})$ per qubit, followed by a CZ entangling ring. Pauli-$Z$ expectations $z=(\langle Z_1\rangle,\dots,\langle Z_{n_q}\rangle)$ feed a small affine head $\tilde u=\mathrm{clip}(W_h z+b_h)$. Trainable parameters are the quantum angles $(\lambda,\theta)$ plus the classical encoder/head; data re-uploading endows $\tilde u(\cdot)$ with a truncated Fourier spectrum whose bandwidth grows with $L$ \citep{schuld2021fourier}, matching the periodic structure of building disturbances.

\paragraph{Intervention budget via primal--dual training.}
Let $I(\theta)=\mathbb{E}\,\|u^{\mathsf P}-\tilde u\|/\mathrm{span}(\mathcal U)$ be the normalized CBF intervention. We pose intervention-aware learning as constrained optimization,
\begin{equation}
\min_\theta\ \mathcal{L}_{\mathrm{task}}(\theta)\quad\text{s.t.}\quad I(\theta)\le B,
\label{eq:constrained}
\end{equation}
and solve its Lagrangian $\mathcal{L}_{\mathrm{task}}+\lambda\,(I-B)$ by projected dual ascent,
\begin{equation}
\lambda_{k+1}=\big[\lambda_k+\eta_\lambda(\hat I_k-B)\big]_+ .
\label{eq:dual}
\end{equation}
Sweeping $B$ traces the reliance--task Pareto front. The task loss combines behavior cloning of the logged safe action, a one-step differentiable comfort loss through $f_\phi$, a learned differentiable energy-head loss, and an action-support (guard) penalty:
\begin{equation}
\mathcal{L}_{\mathrm{task}}=w_{\mathrm{bc}}\,\ell_{\mathrm{bc}}+w_{\mathrm{room}}\,\ell_{\mathrm{room}}+w_{\mathrm{en}}\,\ell_{\mathrm{en}}+w_{\mathrm{ag}}\,\ell_{\mathrm{ag}}.
\label{eq:task}
\end{equation}

\paragraph{Guarded learned-energy head.}
$\ell_{\mathrm{en}}$ uses a learned head predicting $\log(1{+}\text{energy})$ from the same context and action. A learned energy model is differentiable and convenient but can be \emph{exploited out of distribution}: minimizing predicted energy may drive the policy toward actuator combinations the head never saw. We therefore (i) add $\ell_{\mathrm{ag}}$ to keep raw actions inside the logged support during training, and (ii) keep the runtime guard $\mathsf G$ at deployment. Section~\ref{sec:exp} shows both are necessary.

\paragraph{Safety-attribution evaluation and guard-off.}
At evaluation we log the decomposition~\eqref{eq:decomp}, $\Vpre$, and $\Vpost$ for every step, and we run each policy twice: once with the runtime guard \emph{on} (deployment setting) and once \emph{off} ($\mathsf G=\mathrm{id}$). Guard-off isolates whether a method's apparent safety is intrinsic to the policy or supplied by the guard, and whether the learned energy head is safe on its own. Unless stated otherwise, tables report per-episode means of $\Vpre$, $\Vpost$, and the corrections $c_{\bullet}$, averaged over the $60$ matched episodes per method and aggregated across the $5$ seeds.

\begin{algorithm}[t]
\caption{$\Iaqdpc$ training (per testcase)}
\label{alg:main}
\begin{algorithmic}[1]
\STATE \textbf{Input:} logged context/action data, model $f_\phi$, energy head, budget $B$, weights $w_\bullet$, dual rate $\eta_\lambda$
\STATE init VQC params $\theta$, multiplier $\lambda\!\leftarrow\!\lambda_0$
\FOR{epoch $=1,\dots,E$}
  \STATE $\tilde u\!\leftarrow\!\pi_\theta(s)$;\ \ $u^{\mathsf P}\!\leftarrow\!\mathsf P(x,\tilde u,d)$ via \eqref{eq:proj}
  \STATE $\mathcal L\!\leftarrow\!\mathcal L_{\mathrm{task}}(\theta)+\lambda\big(I(\theta)-B\big)$ \hfill // \eqref{eq:task},\eqref{eq:constrained}
  \STATE $\theta\!\leftarrow\!\text{Adam}(\nabla_\theta\mathcal L)$ \hfill \citep{kingma2015adam}
  \STATE $\lambda\!\leftarrow\![\lambda+\eta_\lambda(\hat I-B)]_+$ \hfill // \eqref{eq:dual}
\ENDFOR
\STATE \textbf{return} $\theta$
\end{algorithmic}
\end{algorithm}

\section{Analysis}
\label{sec:theory}
We record two facts that justify the design. Proofs are in the supplement.

\begin{proposition}[Shadow price of the intervention budget]
\label{prop:dual}
For~\eqref{eq:constrained} with continuous $\mathcal{L}_{\mathrm{task}}$ and $I\ge0$, the update~\eqref{eq:dual} is projected dual ascent on the concave dual $g(\lambda)=\min_\theta \mathcal{L}_{\mathrm{task}}+\lambda(I-B)$. Under convexity and a Slater point, $\lambda_k\to\lambda^\star$ and the running primal average is asymptotically feasible with $\mathcal{L}_{\mathrm{task}}$ within $O(1/\sqrt{K})$ of the constrained optimum; moreover $\partial \mathcal{L}^\star_{\mathrm{task}}/\partial B=-\lambda^\star\le 0$.
\end{proposition}
Thus tightening the reliance budget $B$ cannot reduce achievable task cost---exactly the reliance--comfort trade-off we observe.

\begin{proposition}[Minimal, well-defined correction]
\label{prop:proj}
If $\{u\in[\underline u,\overline u]:Gu\ge g\}\neq\varnothing$, the exact-penalty program~\eqref{eq:proj} with $\rho$ large enough returns the unique Euclidean projection of $\tilde u$ onto that set, and $u^{\mathsf P}$ is differentiable in $\tilde u$ almost everywhere. Hence $c_{\mathrm{cbf}}=\|u^{\mathsf P}-\tilde u\|$ is a well-defined, minimal CBF correction.
\end{proposition}

\section{Experiments}
\label{sec:exp}

\paragraph{Setup.}
We use BOPTEST v0.9.0 \citep{blum2021boptest} closed loops at a $15$-minute step ($96$ steps/episode) on two hydronic test cases, \textsc{singlezone\_commercial\_hydronic} and \textsc{twozone\_apartment\_hydronic}. (A third case, \textsc{bestest\_hydronic\_heat\_pump}, is \emph{empirically degenerate}---all methods yield identical trajectories under the logged-safe envelope, confirmed by a guard-off screen---and is excluded from the main table and reported in the supplement.) Each case spans two heating periods and three electricity-price profiles; we run $5$ seeds, giving $60$ episodes per method ($420$ guarded and $300$ guard-off closed-loop episodes total, $0$ errors). We compare a rule-based controller; raw behavior-cloning policies and their intervention-aware counterparts for three policy classes---a large MLP ($\approx\!6.9$k params), an \emph{equal-parameter} MLP (\textsc{MLP}$_s$, $\approx\!408$), and a VQC ($\approx\!400$, of which only $12$ are trainable quantum angles). Significance uses a paired sign-flip permutation test across matched scenarios with Cliff's $\delta$ effect size and $95\%$ bootstrap confidence intervals (CIs); $p$ values report two-sided sign-flip probabilities.

\begin{table}[t]
\centering\small
\setlength{\tabcolsep}{4pt}
\renewcommand{\arraystretch}{1.15}
\begin{tabular}{@{}p{0.45\columnwidth}p{0.47\columnwidth}@{}}
\toprule
\textbf{Claim} & \textbf{Where it is earned} \\
\midrule
Intervention-aware training lowers the quantum policy's raw violation \emph{and} safety-layer reliance & $\Vpre$ and $c_{\mathrm{tot}}$ drop with $p<10^{-4}$ and large Cliff's $\delta$ (Table~\ref{tab:sig}) \\
\dots\ at no significant energy cost & Energy change not significant ($p{=}0.06$, Table~\ref{tab:sig}) \\
The gain is policy-level, not filter-supplied & Guard-off: $\Iaqdpc$ stays safest with the guard removed (Fig.~\ref{fig:go}) \\
Safer \emph{and} more comfortable than a matched classical policy at equal capacity & Equal-${\approx}400$-param comparison (Fig.~\ref{fig:eq}, Table~\ref{tab:main}) \\
A learned energy head is safe \emph{only} with a runtime guard & Guard-off exposes pathological loops without it (Fig.~\ref{fig:go}) \\
\bottomrule
\end{tabular}
\caption{Claim-to-evidence map: every headline claim is tied to a specific significance test or stress test.}
\label{tab:claims}
\end{table}

\begin{table*}[t]
\centering
\small
\setlength{\tabcolsep}{5pt}
\begin{tabular}{lrrrrrrrr}
\toprule
Method & Params & $\Vpre\!\downarrow$ & $\Vpost\!\downarrow$ & $c_{\mathrm{cbf}}$ & $c_{\mathrm{guard}}$ & $c_{\mathrm{tot}}\!\downarrow$ & Comfort\,(K)$\downarrow$ & Energy\,(kWh) \\
\midrule
Rule-based            & $0$    & $0.736$ & $0.302$ & $0.516$ & $0.000$ & $0.516$ & $0.528$ & $499.0$ \\
MLP-raw-clone         & $6.9$k & $0.035$ & $0.050$ & $0.086$ & $0.504$ & $0.590$ & $\mathbf{0.100}$ & $616.8$ \\
IA-MLP-DPC            & $6.9$k & $0.028$ & $0.159$ & $0.089$ & $0.140$ & $0.229$ & $0.236$ & $586.4$ \\
MLP$_s$-raw-clone     & $408$  & $0.029$ & $0.050$ & $0.129$ & $0.340$ & $0.469$ & $\mathbf{0.100}$ & $616.2$ \\
IA-MLP$_s$ (eq.\ param) & $408$ & $0.053$ & $0.363$ & $0.121$ & $0.138$ & $0.259$ & $0.591$ & $560.9$ \\
VQC-raw-clone         & $400$  & $0.048$ & $0.316$ & $0.140$ & $0.189$ & $0.329$ & $0.525$ & $537.9$ \\
\textbf{IA-VQC-DPC (ours)} & $400$ & $0.043$ & $0.305$ & $0.138$ & $\mathbf{0.123}$ & $\mathbf{0.261}$ & $0.514$ & $588.7$ \\
\bottomrule
\end{tabular}
\caption{Guarded closed-loop results pooled over both test cases ($5$ seeds, $n{=}60$ episodes/method). Lower is better for violations and correction. The total correction $c_{\mathrm{tot}}=c_{\mathrm{cbf}}{+}c_{\mathrm{guard}}$ (Eq.~\ref{eq:decomp}) reveals that raw clones rely overwhelmingly on the \emph{runtime guard} (e.g.\ MLP-raw $85\%$), whereas IA-VQC-DPC attains the lowest reliance among learned policies.}
\label{tab:main}
\end{table*}

\begin{figure*}[t]
\centering
\includegraphics[width=0.92\textwidth]{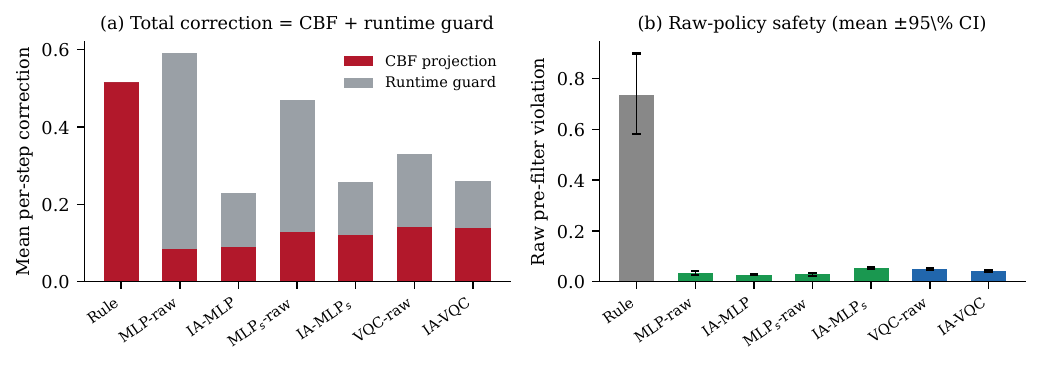}
\caption{Safety attribution on guarded closed loops ($n{=}60$). (a) Total correction decomposes into CBF and runtime-guard parts: the headline ``correction reduction'' of raw clones is mostly deployment guard, not learned safety. (b) Raw pre-filter violation with $95\%$ CIs: only the rule-based controller is grossly unsafe pre-filter; among learned policies, intervention-aware training keeps the compact VQC's raw violations low.}
\label{fig:attr}
\end{figure*}

\paragraph{Finding 1: filters mask, and reliance is confounded.}
Table~\ref{tab:main} and Fig.~\ref{fig:attr} make the masking concrete. The rule-based controller is grossly unsafe \emph{pre}-filter ($\Vpre{=}0.736$) yet its \emph{post}-filter loop looks ordinary ($\Vpost{=}0.302$): the filter earns its safety. For learned policies the dominant correction is the \emph{runtime guard}, not the CBF---e.g.\ MLP-raw-clone has $c_{\mathrm{guard}}/c_{\mathrm{tot}}{=}85\%$. A naive ``correction-norm'' comparison would therefore credit the deployment guard to the method; our decomposition~\eqref{eq:decomp} prevents this and is, to our knowledge, the first closed-loop safety attribution of this kind.

\paragraph{Finding 2: intervention-aware training lets the quantum policy earn its safety.}
Relative to its own raw clone, IA-VQC-DPC significantly lowers raw pre-filter violation ($\Delta\Vpre{=}{-}0.0055$, $95\%$ CI $[-0.0068,-0.0042]$, $\delta{=}{-}0.25$, $p{<}10^{-4}$) and total safety-layer reliance ($\Delta c_{\mathrm{tot}}{=}{-}0.068$, CI $[-0.088,-0.051]$, $\delta{=}{-}0.44$, $p{<}10^{-4}$), with no significant energy regression ($p{=}0.06$) and slightly better comfort ($p{=}0.039$); see Table~\ref{tab:sig}. The classical IA-MLP, in contrast, reduces reliance only by trading comfort ($\Delta\text{comfort}{=}{+}0.136\degK$, $\delta{=}0.63$, $p{<}10^{-4}$): intervention-awareness is most favorable for the compact quantum policy.

\begin{figure}[t]
\centering
\includegraphics[width=0.86\columnwidth]{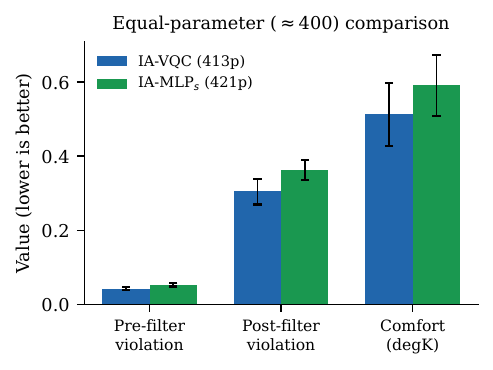}
\caption{Equal-parameter ($\approx\!400$) comparison with $95\%$ CIs. The intervention-aware quantum policy is significantly safer (pre/post-filter) and more comfortable than a matched-capacity classical policy.}
\label{fig:eq}
\end{figure}

\paragraph{Finding 3: at equal capacity, the quantum policy is safer and more comfortable.}
The decisive fairness test compares IA-VQC against the equal-parameter IA-MLP$_s$ ($\approx\!400$ params each). IA-VQC is significantly safer pre- and post-filter ($\Delta\Vpre{=}{-}0.010$, $\Delta\Vpost{=}{-}0.058$, both $p{<}10^{-4}$) and significantly more comfortable ($\Delta\text{comfort}{=}{-}0.077\degK$, $p{<}10^{-4}$), at a significant energy cost ($+27.8$ kWh, $p{=}0.003$). We report this honestly as a safety/comfort--energy Pareto trade-off (Fig.~\ref{fig:eq}): at a fixed, tiny parameter budget the quantum inductive bias buys safety and comfort.

\begin{figure}[t]
\centering
\includegraphics[width=\columnwidth]{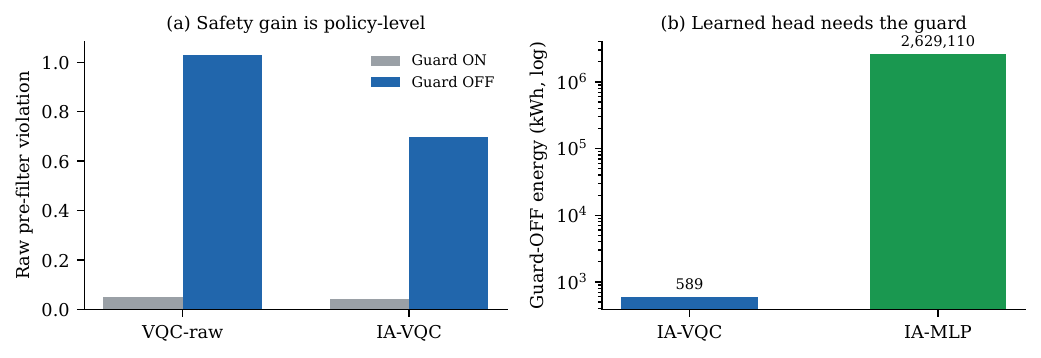}
\caption{Guard-off evaluation. (a) Removing the runtime guard, IA-VQC still has significantly lower raw pre-filter violation than VQC-raw ($\Delta{=}{-}0.33$, $p{<}10^{-4}$): the gain is policy-level. (b) Without the guard, the learned-energy-head IA-MLP is exploited out of distribution and its energy explodes ($2.6\!\times\!10^{6}$ kWh), while IA-VQC remains stable---motivating the guarded design.}
\label{fig:go}
\end{figure}

\paragraph{Finding 4: guard-off isolates policy-level safety and a valuable negative result.}
With the runtime guard removed, IA-VQC's advantage over VQC-raw \emph{strengthens} and the CBF-specific reduction becomes significant ($\Delta\Vpre{=}{-}0.333$, $p{<}10^{-4}$; $\Delta c_{\mathrm{cbf}}{=}{-}0.184$, $p{<}10^{-4}$), confirming the improvement is intrinsic to the policy rather than supplied by the guard. Guard-off also exposes the negative result (Fig.~\ref{fig:go}b): the learned differentiable energy head, when unguarded, drives IA-MLP to physically pathological actuator combinations with energy of $2.6\!\times\!10^{6}$ kWh, whereas the guarded design and the quantum policy remain stable. A learned energy model is useful for control \emph{only} with distribution-aware runtime support.

\begin{table}[t]
\centering
\small
\setlength{\tabcolsep}{3.5pt}
\begin{tabular}{llrrr}
\toprule
Comparison & Metric & $\Delta$ & $\delta$ & $p$ \\
\midrule
\multirow{4}{*}{\shortstack[l]{IA-VQC vs.\\ VQC-raw}}
 & $\Vpre$    & $-0.0055$ & $-0.25$ & $<10^{-4}$ \\
 & $c_{\mathrm{tot}}$ & $-0.068$ & $-0.44$ & $<10^{-4}$ \\
 & Comfort    & $-0.011$ & $-0.11$ & $0.039$ \\
 & Energy     & $+50.8$ & $0.03$ & $0.060$ \\
\midrule
\multirow{4}{*}{\shortstack[l]{IA-VQC vs.\\ IA-MLP$_s$\\ (eq.\ param)}}
 & $\Vpre$    & $-0.010$ & $-0.25$ & $<10^{-4}$ \\
 & $\Vpost$   & $-0.058$ & $-0.22$ & $<10^{-4}$ \\
 & Comfort    & $-0.077$ & $-0.26$ & $<10^{-4}$ \\
 & Energy     & $+27.8$ & $-0.13$ & $0.003$ \\
\bottomrule
\end{tabular}
\caption{Paired sign-flip tests ($n{=}60$): mean difference $\Delta=$mean(A)$-$mean(B), Cliff's $\delta$, and $p$. All $p{<}10^{-4}$ entries have $95\%$ bootstrap CIs of $\Delta$ excluding $0$; full CIs are in the supplement. Negative $\Delta$ favors the first method for violations/correction/comfort.}
\label{tab:sig}
\end{table}

\section{Discussion}
Our results make a sharp, positive case. Intervention-aware training lets a compact quantum policy \emph{earn} its safety: it significantly lowers raw pre-filter violation and total safety-layer reliance, and under guard-off stress---with the protective layers removed---it remains the safest learned policy, so the gain is intrinsic to the policy rather than supplied by a filter. At an equal ${\approx}400$-parameter budget the quantum policy is simultaneously safer and more comfortable than a matched classical network: a Pareto-favorable safety/comfort outcome at genuinely matched capacity. The same evaluation that certifies this earned safety also turns the learned energy head into an actionable design rule: it helps \emph{only} when paired with a distribution-aware runtime guard, which any learned-dynamics controller should adopt. Two scope notes keep the claims precise: the quantum content is $12$ trainable angles whose value we demonstrate at equal capacity (a Pareto-favorable safety/comfort outcome, not energy dominance), and the building instantiation uses one-step differentiable model losses (offline intervention-aware DPC). Neither qualifies the two transferable messages: post-filter safety must be \emph{attributed}, and intervention-aware training plus guard-off evaluation can \emph{certify} that a compact policy earned its safety.

\section{Conclusion}
We reframed safe policy learning around a measurable question---\emph{who earns the safety, the policy or its protective layers?}---and answered it with a safety-attribution evaluation protocol and an intervention-aware variational-quantum differentiable predictive controller. On high-fidelity BOPTEST closed loops, $\Iaqdpc$ significantly reduces a compact quantum policy's raw pre-filter violation and safety-layer reliance, is safer and more comfortable than an equal-parameter classical policy, and---under guard-off evaluation---demonstrably earns its safety while revealing that a learned energy head needs a runtime guard. The attribution protocol is independent of the policy class and the application, and we hope it becomes a standard lens for filtered learning-based control.

\bibliography{references}

\end{document}